*Jia Hu, Xuerun Yan, Tian Xu, Haoran Wang*

# Automated Driving with Evolution Capability: A Reinforcement Learning Method with Monotonic Performance Enhancement

**Jia Hu, Ph.D.**
ZhongTe Distinguished Chair in Cooperative Automation, Professor
Key Laboratory of Road and Traffic Engineering of the Ministry of Education
Tongji University, No.4800 Cao'an Road, Shanghai, China, 201804
Email: hujia@tongji.edu.cn

**Xuerun Yan**
Research Assistant
Key Laboratory of Road and Traffic Engineering of the Ministry of Education
Tongji University, No.4800 Cao'an Road, Shanghai, China, 201804
Email: frankyan@tongji.edu.cn

**Tian Xu**
Research Assistant
Key Laboratory of Road and Traffic Engineering of the Ministry of Education
Tongji University, No.4800 Cao'an Road, Shanghai, China, 201804
Email: xutian_work@163.com

**Haoran Wang, Ph.D., Corresponding Author**
Associate Researcher
Key Laboratory of Road and Traffic Engineering of the Ministry of Education
Tongji University, No.4800 Cao'an Road, Shanghai, China, 201804
Email: wang_haoran@tongji.edu.cn

*Jia Hu, Xuerun Yan, Tian Xu, Haoran Wang*

**ABSTRACT**

Reinforcement Learning (RL) offers a promising solution to enable evolutionary automated driving. However, the conventional RL method is always concerned with risk performance. The updated policy may not obtain a performance enhancement, even leading to performance deterioration. To address this challenge, this research proposes a High Confidence Policy Improvement Reinforcement Learning-based (HCPI-RL) planner. It is intended to achieve the monotonic evolution of automated driving. A novel RL policy update paradigm is designed to enable the newly learned policy's performance consistently surpass that of previous policies, which is deemed as monotonic performance enhancement. Hence, the proposed HCPI-RL planner has the following features: i) Evolutionary automated driving with monotonic performance enhancement; ii) With the capability of handling scenarios with emergency; iii) With enhanced decision-making optimality. Results demonstrate that the proposed HCPI-RL planner enhances the policy return by 44.7% in emergent cut-in scenarios, 108.2% in emergent braking scenarios, and 64.4% in daily cruising scenarios, compared to the PPO planner. Adopting the proposed planner, automated driving efficiency is enhanced by 19.2% compared to the PPO planner, and by 30.7% compared to the rule-based planner.

**Keywords:** Automated driving, Reinforcement Learning, Motion Planning, Policy Improvement, Monotonic Performance Enhancement

*Jia Hu, Xuerun Yan, Tian Xu, Haoran Wang*

## 1 Introduction

With the maturation of cutting-edge technologies such as Artificial Intelligence (AI) and the Internet of Things (IoT), automated driving technology is undergoing rapid development (Van Brummelen et al., 2018; Zhang et al., 2023a). However, widespread commercialization of automated driving in society still remains a distant goal (He et al., 2024b; Kalra and Paddock, 2016). Automated driving is challenged for effectively navigating in highly dynamic, complex, or even emergent scenarios (Huang et al., 2023). To be adaptive to these scenarios, the automated driving system should ensure performance enhancement when there is more driving data collected. Hence, the evolution capabilities are quite crucial for automated driving (Du et al., 2024).

Reinforcement Learning (RL) offers a promising solution to enable evolutionary automated driving (Wu et al., 2024; Yuan et al., 2024). The RL method updates the control policy by interacting with the environment (Sutton and Barto, 2018). By reinforcing positive actions and adjusting negative ones, RL refines the driving policy and becomes more proficient at handling various scenarios (Elallid et al., 2022; He et al., 2024a). Previous studies have demonstrated the effectiveness of RL-based automated driving in specific scenarios (Aradi, 2020; Kiran et al., 2021). For instance, RL has successfully learned to maintain desired speed in highway driving (Changxi You, 2019; Nageshrao et al., 2019), navigate intersections with varying traffic conditions (Isele et al., 2018; Yang et al., 2024), and execute on-ramp merging maneuvers seamlessly (Chen et al., 2023; Lin et al., 2022). These applications highlight RL's capability to enhance the performance of automated driving systems.

However, concerns persist regarding the monotonic evolution of the conventional RL algorithm (Metelli et al., 2021; Pirotta et al., 2013). The RL-based automated vehicle explores new experiences in the interaction with surrounding vehicles. Based on the newly explored experience, a conventional RL algorithm would update policy without the consideration of old experience (Li, 2023). The exploration may not always find a better action. Thereby, the updated policy may not obtain a performance enhancement, even leading to performance deterioration (Cao et al., 2023; Laroche et al., 2019). For example, as shown in **Fig. 1(b)**, the automated vehicle drives safely with the old policy, the policy in the blue point of **Fig. 1(a)**. When the RL algorithm explores new strategies, it may update a policy with less return, as shown by the orange point in **Fig. 1(a)**. Using the new policy, the vehicle collides with background vehicles, as shown in **Fig. 1(c)**. This indicates that for the conventional RL algorithm, with further training, the newly learned policy might perform worse than the previous policy. Exploration and policy update may jeopardize driving safety and efficiency (Thomas and Brunskill, 2016; Thomas et al., 2015b). We never know if a RL-based planner would fail in the next second. This is a significant reason why RL-based automated driving has not been commercialized. Therefore, it is crucial to ensure a monotonic performance enhancement of RL-based automated driving.

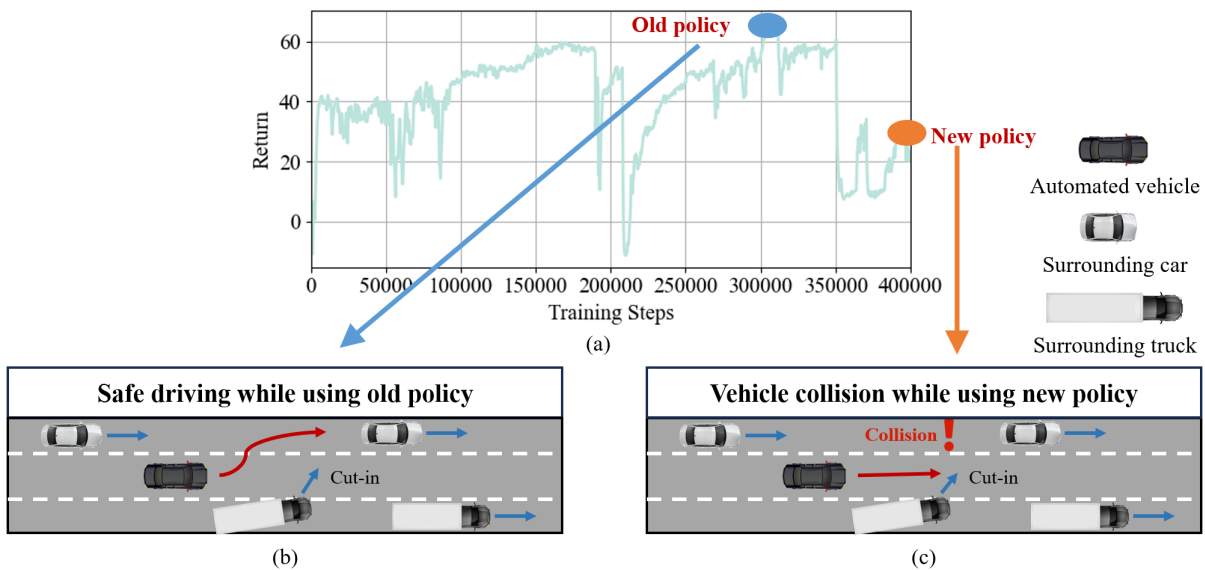

**Fig. 1.** Illustration of the evolution reliability concern about RL-based automated driving. (a) RL-based automated driving performance; (b) automated vehicle drives safely while using old policy; (c) automated vehicle has a collision while using new policy.

Current studies on the performance enhancement of RL-based automated driving still have limitations. These studies are primarily categorized into three types: expert-augmented learning, incorporating safety checkers and robust RL. i) Expert-augmented learning, e.g., apprenticeship learning (Abbeel and Ng, 2004) and imitation learning (Le Mero et al., 2022), aims to mimic an expert's behavior. It integrates human experts' demonstrations with the RL framework, and thus leverages expert knowledge to improve the RL performance (Zhu and Zhao, 2021). Hence, this approach cannot surpass the expert's policy. Even worse, it will fail in scenarios that the expert has never encountered before. ii) Incorporating the safety checkers with RL is a method to create agents that not only learn effective policies but also operate with safety boundaries (Baheri et al., 2020; Wachi and Sui, 2020). The safety checkers evaluate the risk of RL action and modify the unsafe action. For example, some researchers adopted rule-based policy (Cao et al., 2022; Cao et al., 2021; Yang et al., 2024) or expert knowledge (Li et al., 2022) to guarantee a lower-bound performance. Although this approach ensures the safety of RL policy update, it cannot ensure a monotonic performance enhancement in each update. iii) Robust RL aims to maintain safety and high performance in the face of uncertainty and risk (Wu et al., 2024; Yang et al., 2023). It optimizes the RL policy in an uncertain or even adversarial environment. For example, a defense-aware robust RL has been proposed to enhance the automated driving performance against worst-case observational perturbations (He et al., 2024b). However, in this study, the policy performance exhibits fluctuations over training time and still fails to achieve monotonic enhancement with policy updates. In summary, all the aforementioned approaches still cannot enable the monotonic performance enhancement of automated driving.

Moreover, most RL-based automated driving studies consider behavioral decision-making and motion planning separately. They typically adopted a hierarchical framework (Lu et al., 2023; Pateria et al., 2021). For instance, at the high-level of decision-making, RL methods are employed to make behavioral decisions(Gu et al., 2023), such as lane keeping, lane changing or turning. At the low-level, Model Predictive Control (MPC) methods are adopted to output motion commands (Al-Sharman et al., 2023). Such a hierarchical framework cannot ensure a globally optimal solution. Specifically, the behavior decision does not consider the feasibility of the subsequent motion plan, resulting in suboptimal or even infeasible maneuvers (Li et al., 2023). Low-level motion plan may not accurately execute the planned behavior, since the high-level behavior planning does not consider detailed vehicle dynamics which constrains execution capability (Zhang et al., 2023b). Additionally, separating behavioral decision-making and motion planning limits the flexibility to adaptively respond to dynamic changes in the environment. The motion planning has to work within the constraints of the predetermined behavior, delaying the necessary adaptation to ensure safety and efficiency. Therefore, integrating decision-making and motion planning is essential for achieving enhanced optimal maneuvers.

Considering the aforementioned shortcomings, this research proposes a novel RL-based planner that ensures monotonic performance enhancement of automated driving. It has the following features:

- Evolutionary automated driving with monotonic performance enhancement;
- With the capability of handling scenarios with emergency;
- With enhanced decision-making optimality.

The remainder of the paper is organized as follows: Section 'Methodology' proposes the framework and introduces the detailed formulation of the proposed planner; Section 'Performance evaluation' evaluates the proposed planner in terms of emergent scenarios and daily cruising scenarios; Section 'Conclusion' makes a conclusion and discusses the future study.

## 2  Methodology

The goal of the paper is to propose an RL-based planner that enables monotonic performance enhancement. This new planner is named by High Confidence Policy Improvement RL-based (HCPI-RL) planner. The rationale of this planner is to design a new policy improvement paradigm that provides a high confidence guarantee about performance enhancement.

*Jia Hu, Xuerun Yan, Tian Xu, Haoran Wang*

## 2.1 Highlights

There are three highlights of the proposed planner:

- **Monotonic performance enhancement by high confidence policy improvement**: The proposed planner combines the RL method with a high confidence policy improvement scheme. The newly learned policy's performance is evaluated based on the previously collected dataset. The proposed planner only allows the policy update when the newly learned policy surpasses the previous policy with high confidence. In this way, the driving policy achieves incremental improvements with each update, leading to the monotonic evolution of automated driving.
- **With lower-bound performance ensured**: Besides the monotonic performance enhancement, a rule-based policy is designed as a safety checker for the proposed HCPI-RL planner. The RL policy is implemented only when it surpasses the rule-based policy. It prevents the unsafe implementation of RL policy, especially at the early training stage.
- **Integrate both decision-making and motion planning**: The proposed planner integrates both behavior decision-making and motion planning to achieve a global optimal maneuver. It directly outputs the optimized driving commands based on the environment information input. Such an integrated design considers the execution capability of the subsequent motion plan in the behavioral decision-making phase. In this way, the integrated planner is able to not only timely response to environmental changes, but also get an enhanced optimal maneuver.

## 2.2 System framework

The system framework of the HCPI-RL planner is illustrated in **Fig. 2**. It consists of three modules. Details of each module are as follows:

1) **Module I**: This module gets and collects the information of ego vehicle and surrounding vehicles. It consists of an environment and replay buffer. The environment represents the external world where an agent can interact and learn. It inputs the vehicle actions and outputs the vehicle states and action rewards. The replay buffer is to store and reuse past trajectories. It allows the agent learns from past experiences to improve learning efficiency and stability. The replay buffer is split into a training dataset and a test dataset.
2) **Module II**: This module makes the motion command. It consists of an RL-based generator, a confidence discriminator, and a hybrid policy application component. i) The RL-based generator adopts an Actor-Critic (AC) structure. It generates the candidate policy that can maximize the future reward via trial and error with the environment. ii) The confidence discriminator discriminates whether the candidate policy is of high confidence for performance enhancement. It includes two components: high confidence policy evaluation and high confidence policy improvement. High confidence policy evaluation assesses the lower-bound performance of the candidate policy. High confidence policy improvement updates the policy if the lower-bound performance of the candidate policy surpasses the current policy performance. The policy update means that the candidate policy outperforms the current policy with high confidence. iii) The hybrid policy application is to choose the better one from the RL policy and rule-based policy. The rule-based policy is designed as a safety checker to avoid bad performance of RL policy especially at the early-training stage.
3) **Module III**: This module receives the driving motion commands from the proposed planner and actuates the ego vehicle.

*Jia Hu, Xuerun Yan, Tian Xu, Haoran Wang*

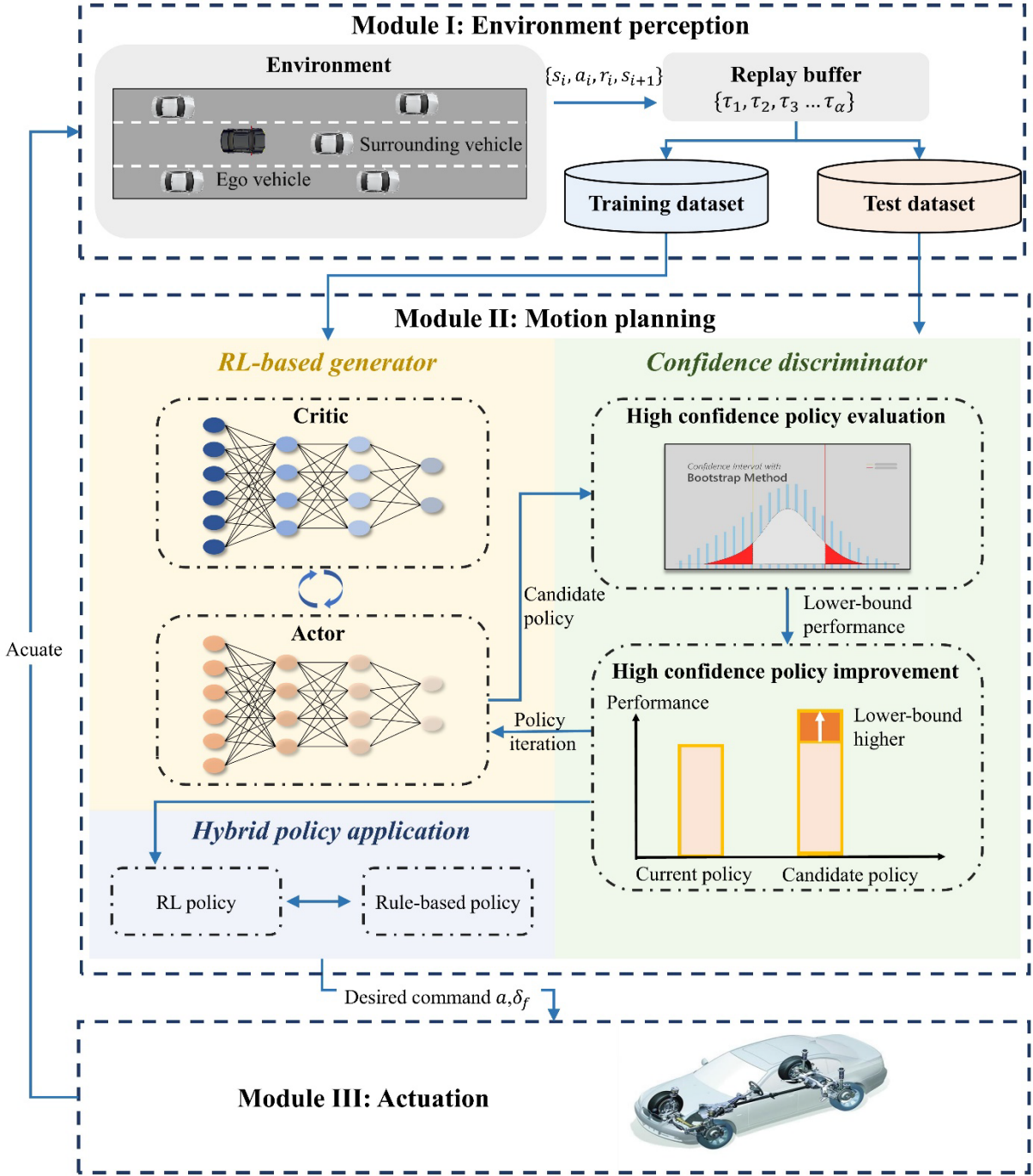

**Fig. 2.** The system framework of the HCPI-RL planner.

## 2.3 Problem definition

The proposed HCPI-RL planner is adopted for the motion planning of an automated vehicle. The RL-based motion planner should evolve to adapt dynamic environment. When updating policy with new data, conventional RL methods may result in a new policy with worse performance, potentially leading to serious safety issues. To ensure monotonic performance enhancement, this paper proposes the HCPI-RL planner with high confidence policy improvement.

The basis of the proposed method is a Constrained Markov Decision Process (CMDP) problem (Altman, 2021). The CMDP is an augmented MDP with constraints. The CMDP is defined by a tuple $\langle S, A, P, r, \gamma \rangle$. $S$ is the state space. $A$ is the action space. $P: S \times S \times A \to \mathbb{R}$ is the transition probability distribution. $r: S \times A \to \mathbb{R}$ is the reward function. $\gamma \epsilon (0,1]$, is the discounter rate. The policy $\pi(a|s)$ is the function mapping from the state space to action space, which determines the action probability distribution at the given state $s$. The goal of the CMDP is to find an optimal policy that maximizes the future rewards with the constraints.

For the proposed HCPI-RL planner, the goal is to design a policy update paradigm that ensures the new policy satisfies the constraints of a high confidence guarantee of performance enhancement. Hence, the HCIL-RL aims at finding the optimal solution to the following problem:

$$\max_{\pi^*} \mathbb{E}_{\pi^*} \left[ \sum_{t=0}^{\infty} \gamma^t r(s_t, a_t) \right]$$

$$s.t. \, \rho_-(\pi^*) \geq \rho(\pi_{cur})$$

(1)

where $\mathbb{E}$ is the expectation operator, $\pi^*$ and $\pi_{cur}$ are new RL policy and current RL policy, $s_t$, $a_t$ are the state and action at time step $t$ respectively, $\rho_-(\pi^*)$ is the lower-bound confidence value of the new policy performance, $\rho(\pi_{cur})$ is the estimation of the current policy performance. both $\rho_-$ and $\rho$ are defined in the following *section 2.5.1*.

In the proposed HCPI-RL planner, a policy update paradigm is designed to address the above CMDP problem. The paradigm combines an RL-based generator and a confidence discriminator. The detailed structure of the paradigm is shown in **Fig. 3**. The RL-based generator aims to update the policy network and generate a candidate policy. The confidence discriminator includes high confidence policy evaluation and high confidence policy improvement. These two processes are used in combination to ensure the new policy meets the high confidence guarantee of performance enhancement.

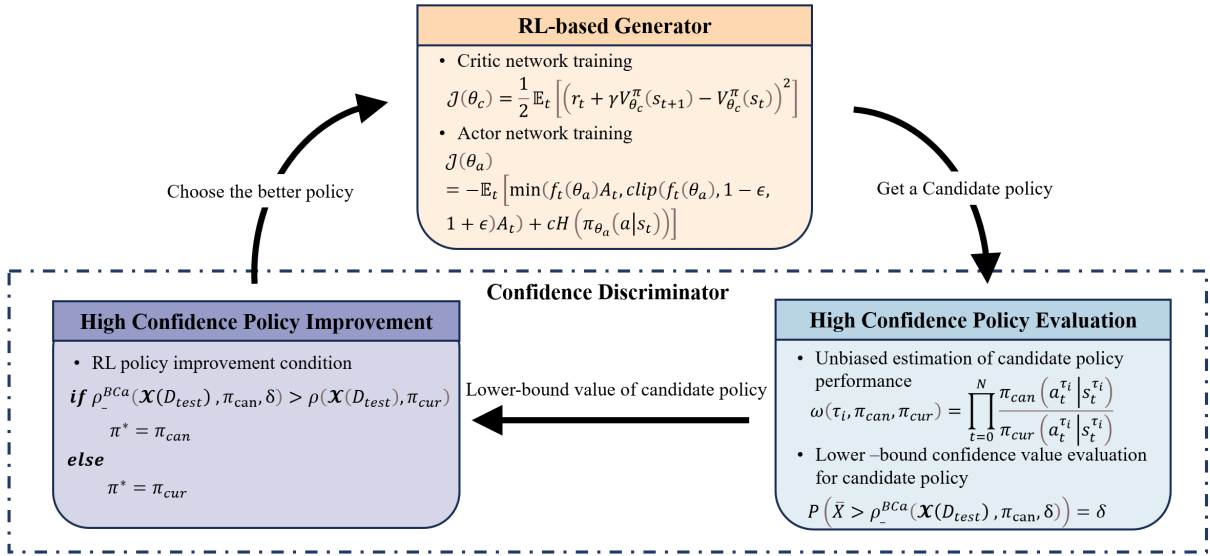

**Fig. 3.** The paradigm combining an RL-based generator and a confidence discriminator.

## 2.4 RL-based generator

An AC structure is utilized in the RL-based generator. The AC structure combines the benefits of both policy-based and value-based methods (Konda and Tsitsiklis, 1999). It consists of two components: an actor and a critic. The actor represents the policy $\pi(a|s)$. It is responsible for formulating and updating the policy. The actor aids the agent in selecting actions that optimize long-term rewards. The critic is responsible for estimating the value function $V^\pi(s)$ for a given policy. The value function is defined in **Eq. (2)**. With the value function estimation, the critic can evaluate the actor's policy and provide feedback to refine the actor's policy.

*Jia Hu, Xuerun Yan, Tian Xu, Haoran Wang*

$$V^\pi(s) = \mathbb{E}_\pi \left[ \sum_{t=0}^{\infty} \gamma^t r_t \, | s_0 = s \right], \forall s \in S \tag{2}$$

Utilize the network with parameter $\theta_c$ to represent the critic. The critic network parameters can be learned by minimizing the following loss functions:

$$\mathcal{J}(\theta_c) = \frac{1}{2} \mathbb{E}_t \left[ \left( r_t + \gamma V_{\theta_c}^\pi(s_{t+1}) - V_{\theta_c}^\pi(s_t) \right)^2 \right] \tag{3}$$

where the loss function is constructed by the Temporal Difference (TD) error, $\left( r_t + \gamma V_{\theta_c}^\pi(s_{t+1}) \right)$ is the target value, $V_{\theta_c}^\pi(s_t)$ is the current value at the state $s_t$, $r_t$ and $s_t$ are the reward and state at time step $t$.

Utilize the network with parameter $\theta_a$ to represent the actor. The actor network parameters can be learned via minimizing the following loss function:

$$\mathcal{J}(\theta_a) = -\mathbb{E}_t \left[ min(f_t(\theta_a) A_t, clip(f_t(\theta_a), 1 - \epsilon, 1 + \epsilon) A_t) + cH\left( \pi_{\theta_a}(a|s_t) \right) \right] \tag{4}$$

with

$$f_t(\theta_a) = \frac{\pi_{\theta_a}(a_t|s_t)}{\pi_{cur}(a_t|s_t)} \tag{5}$$

where $t$ is the time step, $f_t(\theta_a)$ is the probability ratio of policy $\pi_{\theta_a}$ to current policy $\pi_{cur}$, $A_t$ is the estimation of the advantage function, $c$ is the coefficient, $H\left( \pi_{\theta_a}(a|s_t) \right)$ denotes the entropy bonus, the *clip* function constrains the ratio $f_t(\theta_a)$ within the range $[1 - \epsilon, 1 + \epsilon]$, and $\epsilon$ is the hyperparameter for controlling the clipping range. By taking the minimize of unclipped value $f_t(\theta_a) A_t$ and clipped value $clip(f_t(\theta_a), 1 - \epsilon, 1 + \epsilon) A_t$, optimize the loss function can avoid taking an excessively large policy update and keep the stability of training. The $A_t$ and $H\left( \pi_{\theta_a}(a|s_t) \right)$ are defined as follows:

$$A_t = r_t + \gamma r_{t+1} + \cdots + \gamma^{T-1} r_{t+T-1} + \gamma^T V_{\theta_c}^\pi(s_{t+T}) - V_{\theta_c}^\pi(s_t) \tag{6}$$

$$H\left( \pi_{\theta_a}(a|s_t) \right) = -\int \pi_{\theta_a}(a|s_t) \, log \, \pi_{\theta_a}(a|s_t) \, da \tag{7}$$

where $A_t$ quantifies how much better or worse a particular action is compared to the average action taken under a given policy, $T$ is the looking forward steps, $V_{\theta_c}^\pi(s_t)$ and $V_{\theta_c}^\pi(s_{t+T})$ are value functions estimated by the critic at the state $s_t$ and $s_{t+T}$, respectively, $H\left( \pi_{\theta_a}(a|s_t) \right)$ is the function to encourage agent exploration.

Before the actor and critic network training, the RL agent interacts with the environment using the current policy and collects the trajectories to form the replay buffer. The reply buffer allows the agent to reuse past experiences multiple times, making the training process more data-efficient. The replay buffer is defined as follows:

$$\mathcal{B}_{\pi_{cur}} = \{\tau_1, \tau_2, \dots, \tau_\alpha, \} \tag{8}$$

where $\tau$: $\{s_0^\tau, a_0^\tau, r_0^\tau, s_1^\tau, a_1^\tau, r_1^\tau, \dots, s_N^\tau, a_N^\tau, r_N^\tau\}$ is the collected trajectory with the length $N$.

After collecting the trajectory data into the replay buffer, the RL agent randomly samples the tuple $(s_t, a_t, r_t, s_{t+1})$ with the batch size $N_b$ from the reply buffer. The agent constructs the loss functions **Eqs. (3), (4)** with sampled tuples and optimizes the loss with gradient descent. After $K$ epochs of network iterations, the RL agent generates the candidate policy $\pi_{can}$. Each network iteration using the gradient descent is defines as follows:

$$\theta_a^{k+1} = \theta_a^k - \alpha_a \nabla_{\theta_a} \mathcal{J}(\theta_a) \tag{9}$$

$$\theta_c^{k+1} = \theta_c^k - \alpha_c \nabla_{\theta_c} \mathcal{J}(\theta_c) \tag{10}$$

where $k$ in the index of iteration times, $\alpha_a$ and $\alpha_c$ are learning rates for actor and critic network update.

## 2.5 Confidence discriminator

Confidence discriminator is crucial to improve policy performance with a high confidence guarantee. It consists of two components: high confidence policy evaluation and high confidence policy improvement.

### 2.5.1 High confidence policy evaluation

High confidence policy evaluation is to assess the high confidence lower-bound of the candidate policy performance (Thomas et al., 2015a, b). The high confidence lower-bound represents the performance threshold that the candidate policy can surpass with a high confidence. The evaluation method involves inputting the reply buffer data generated by the current policy, and using them to lower-bound the performance of the candidate policy. It includes two steps: i) Using importance sampling (Tokdar and Kass, 2010) to estimate the performance of the candidate policy; ii) Employing Bias Corrected and accelerated (BCa) bootstrap method to determine the high confidence lower-bound.

Importance sampling yields an estimator of the candidate policy performance from the trajectory $\tau_i$, which is generated by the current policy. The estimator is called as importance-weighted return and is defined as follows:

$$\hat{\xi}(\pi_{can}|\tau_i, \pi_{cur}) = \omega(\tau_i, \pi_{can}, \pi_{cur})R(\tau_i) \tag{11}$$

where $\hat{\xi}(\pi_{can}|\tau_i, \pi_{cur})$ is the importance-weighted return, $\pi_{can}$ and $\pi_{cur}$ are candidate policy and current policy respectively, $\omega(\tau_i, \pi_{can}, \pi_{cur})$ is importance weight, $R(\tau_i)$ is the normalized cumulative reward of the trajectory $\tau_i$, and they are defined as follows:

$$\omega(\tau_i, \pi_{can}, \pi_{cur}) = \prod_{t=0}^{N} \frac{\pi_{can}(a_t^{\tau_i}|s_t^{\tau_i})}{\pi_{cur}(a_t^{\tau_i}|s_t^{\tau_i})} \tag{12}$$

$$R(\tau_i) = 2\frac{\sum_{t=0}^{N} \Upsilon^t r(s_t^{\tau_i}) - R^-}{R^+ - R^-} - 1 \tag{13}$$

where $\pi_{can}(a_t^{\tau_i}|s_t^{\tau_i})$, $\pi_{cur}(a_t^{\tau_i}|s_t^{\tau_i})$ are the probability of choosing an action $a_t^{\tau_i}$ at the state $s_t^{\tau_i}$ when using the policy $\pi_{can}$ and $\pi_{cur}$ respectively, $R(\tau_i) \in [-1,1]$, $R^+$, $R^-$ are upper and lower bounds of trajectory reward, and $\Upsilon$ is a discounted rate.

For generality, denote the importance-weighted return $\hat{\xi}(\pi_{can}|\tau_i, \pi_{cur})$ as a random variable $X_i$ and define the test dataset $D_{test} = \{\tau_1, \tau_2, ..., \tau_n\}$ containing $n$ trajectories generated by the current policy. Using the test dataset $D_{test}$, we obtain a set of $n$ random variables $\mathcal{X}(D_{test}) = [X_1, X_2, ..., X_n]$. All these random variables have the same expected value and are the unbiased estimators of the candidate policy performance. Here is the proof:

**Theorem 1**: *Given the set of $n$ random variables $\mathcal{X}(D_{test}) = [X_1, X_2, ..., X_n]$ with the $X_i = \hat{\xi}(\pi_{can}|\tau_i, \pi_{cur})$, all these random variables have the same expected value and are the unbiased estimators of the candidate policy performance.*

**Proof**: Define the following variables:

$$p(\tau|\pi_{can}) = \prod_{t=0}^{N} \pi_{can}(a_t^\tau|s_t^\tau) \tag{14}$$

$$p(\tau|\pi_{cur}) = \prod_{t=0}^{N} \pi_{cur}(a_t^\tau|s_t^\tau) \tag{15}$$

where $p(\tau|\pi_{can})$, $p(\tau|\pi_{cur})$ is the probability of choosing a trajectory $\tau$ when using the candidate policy and current policy, respectively.

Define the candidate policy performance $\rho(\pi_{can})$ as follows:

$$\rho(\pi_{can}) = \mathbb{E}_{\tau \sim \pi_{can}}[R(\tau)] \tag{16}$$

The expected value of each random variable $X_i$ from the set $\mathcal{X}(D_{test})$ is as follows:

$$
\begin{aligned}
\mathbb{E}_{\tau_i \sim \pi_{cur}}[X_i] &= \mathbb{E}_{\tau_i \sim \pi_{cur}}\left[\frac{p(\tau_i|\pi_{can})}{p(\tau_i|\pi_{cur})}R(\tau_i)\right] \\
&= \int p(\tau_i|\pi_{cur})\frac{p(\tau_i|\pi_{can})}{p(\tau_i|\pi_{cur})}R(\tau_i)\,d\tau_i \\
&= \int p(\tau_i|\pi_{can})R(\tau_i)\,d\tau_i \\
&= \mathbb{E}_{\tau \sim \pi_{can}}[R(\tau)]
\end{aligned}
\tag{17}
$$

Therefore,

$$\mathbb{E}_{\tau_i \sim \pi_{cur}}[X_i] = \rho(\pi_{can}) \tag{18}$$

According to **Eq. (18)**, all the random variables $X_i$ are the unbiased estimators of the candidate policy performance, and have the same expected value $\rho(\pi_{can})$. **Theorem 1** is proved.

A BCa bootstrap method ([Jung et al., 2019](#)) is adopted to estimate the high confidence value. It takes a set of importance-weighted returns $\mathcal{X}(D_{test})$ and use it to lower-bound the performance of the candidate policy. The BCa bootstrap method is able to estimate the true distribution of $X_i$ without a specific distribution assumption. The algorithm of BCa bootstrap is defined in **Algorithm 1.**

Define the $\rho_{-}^{BCa}(\mathcal{X}(D_{test}), \pi_{can}, \delta)$ is an estimation of the 1-$\delta$ lower-bound performance of candidate policy from the set $\mathcal{X}(D_{test})$. The performance of candidate policy can surpass the lower-bound with a $\delta$ level of confidence, and is defined as follows:

$$P\left(\bar{X} > \rho_{-}^{BCa}(\mathcal{X}(D_{test}), \pi_{can}, \delta)\right) = \delta \tag{19}$$

where sample mean $\bar{X} = \frac{1}{n}\sum_{i=1}^{n}X_i$ is an unbiased estimator of $\rho(\pi_{can})$.

---

**Algorithm 1 Bias Corrected and accelerated (BCa) bootstrap**

---

**Input**: the confidence level $\delta$, the set $\mathcal{X}(D_{test}) = [X_1, X_2, \ldots, X_n]$ with $X_i = \hat{\xi}(\pi_{can}|\tau_i, \pi_{cur})$, number of bootstrap samples $B$

**Output**: the 1-$\delta$ lower-bound performance of candidate policy $\rho_{-}^{BCa}(\mathcal{X}(D_{test}), \pi_{can}, \delta)$.

1: Compute the sample mean: $\bar{X} = \frac{1}{n}\sum_{i=1}^{n}X_i$.

2: **for** $i = 1$ to $B$ **do**

3:    Randomly sample $n$ elements from the set $\mathcal{X}(D_{test})$, and generate a bootstrap sample $\mathcal{S}_i$.

4:    Compute the mean $\rho_i$ of the bootstrap sample $\mathcal{S}_i$.

5: **end for**

6: Compute the proportion of $\rho_i$ less than $\bar{X}$: $\mathcal{P} = \frac{\sum_{i=0}^{B}N(\rho_i, \bar{X})}{B}$, where $N(\rho_i, \bar{X}) = (\rho_i < \bar{X})?\,1:0$.

7: Compute the bias correction constant $z_0$ via the inverse standard normal CDF: $z_0 = \Phi^{-1}(\mathcal{P})$.

8: **for** $i = 1$ to $n$ **do**

9:    set $y_i$ to be the mean of $\mathcal{X}(D_{test})$ with $i^{th}$ element removed: $y_i = \frac{1}{n-1}\sum_{j=1}^{n}\mathbf{1}_{j \neq i}X_j$, where     $\mathbf{1}_{j \neq i} = (j = i)?\,0:1$.

10: **end for**

11: Compute the mean of $y_i$: $\bar{y} = \frac{1}{n}\sum_{i=1}^{n}y_i$.

12: Compute the acceleration constant $a$: $a = \frac{\sum_{i=1}^{n}(\bar{y}-y_i)^3}{6[\sum_{i=1}^{n}(\bar{y}-y_i)^2]^{3/2}}$.

13: Compute the adjusted percentiles via standard normal CDF: $\beta = \Phi\left(z_0 + \frac{z_0+\Phi^{-1}(1-\delta)}{1-a(z_0+\Phi^{-1}(1-\delta))}\right)$.

14: Determine the BCa confidence lower-bound: $\rho_{-}^{BCa}(\mathcal{X}(D_{test}), \pi_{can}, \delta) = percentile(\boldsymbol{\rho}, 100\beta)$, where $\boldsymbol{\rho}: \{\rho_1, \rho_2, \ldots, \rho_B\}$ is the set of the bootstrap samples' mean.

### 2.5.2 High confidence policy improvement

The policy improvement is to generate a new policy that can be no worse than the current policy performance. The policy improvement takes the current policy performance as the baseline. Once the high confidence lower-bound of the candidate policy exceeds the baseline performance, the RL agent adopts the candidate policy as an improved policy. Otherwise, the RL agent maintains the current policy. The policy improvement process is defined as follows:

$$\pi^* = \begin{cases} \pi_{can}, & if \ \rho_{-}^{BCa}(\mathcal{X}(D_{test}), \pi_{can}, \delta) > \rho(\mathcal{X}(D_{test}), \pi_{cur}) \\ \pi_{cur}, & else \end{cases}$$ (20)

where $\rho^{BCa}$ and $\rho$ are both estimated from the test dataset $D_{test}$. Based on **Eq. (20)**, the RL agent replaces the current policy with the candidate policy when the confidence of the candidate policy is sufficiently high. Therefore, the policy performance enhancement can be guaranteed with high confidence. Upon adoption of the candidate policy, it becomes the current policy in the next iteration, allowing the RL agent to generate a new candidate policy. As a result, through successive policy iterations, the policy performance is expected to enhance monotonically.

## 2.6 Technical implementation

This section provides the technical implementation details of the proposed HCPI-RL planner, including algorithm design, state and action space, reward function, network and hyperparameter.

### 2.6.1 Algorithm

**Algorithm 2** presents a detailed, step-by-step outline of the proposed HCPI-RL method for evolutionary automated driving. The parameters of both the actor and critic networks are initialized using a random distribution. For data collection, the agent interacts with the environment using the current policy $\pi_{cur}$, and saves the collected trajectories into the replay buffer. For network training, the parameters of the actor and critic networks are optimized by combining **Eqs. (3), (4), (9), (10)**. For high confidence policy evaluation, the importance-weighted return of the candidate policy is calculated via **Eq. (11)** and the lower-bound performance is computed via **Algorithm 1**. The agent then selects the better policy using **Eq. (20)**. Once the candidate policy outperforms the current policy with high confidence, the policy is enhanced by adopting the candidate policy. In this way, incremental policy improvement is achieved through successive policy iterations.

---

**Algorithm 2 High Confidence Policy Improvement RL (HCPI-RL)**

---

**Input**: actor network $\theta_a$, critic network $\theta_c$, replay buffer $\mathcal{B}_{\pi_{cur}}$, training dataset $D_{train}$, test dataset $D_{test}$, confidence level $\delta$, maximum training step $M$, maximum episode step $N$, number of collected trajectories $\beta$, epochs $K$, batch size $N_b$, learning rate $\alpha_a, \alpha_c$.

**Output**: RL policy $\pi^*$

1: Initialize the actor network $\theta_a$ and critic network $\theta_c$ with a random distribution, initialize the current policy: $\pi_{cur} \leftarrow \pi_{\theta_a}$.

2: $\mathcal{B}_{\pi_{cur}} \leftarrow \emptyset, D_{train} \leftarrow \emptyset, D_{test} \leftarrow \emptyset, m \leftarrow 0$.

3: **while** $m \leq M$ **do**

4:    **for** i = 1 to $\beta$ **do**

5:        Run the current policy $\pi_{cur}$ in the environment for $N$ steps and generate a trajectory $\tau_i$

6:        Compute the advantage estimate $A_1, \ldots, A_N$ via **Eq. (6)**.

7:    **end for**

8:    Save the collected trajectory into the replay buffer: $\mathcal{B}_{\pi_{cur}} \leftarrow \mathcal{B}_{\pi_{cur}} \cup \{\tau_1, \ldots, \tau_\beta\}$.

9:    Append $\beta/3$ trajectories into $D_{train}$ and the rest into $D_{test}$: $D_{train} \leftarrow D_{train} \cup \{\tau_1, \tau_4, \ldots, \tau_{3n+1}\}$, $D_{test} \leftarrow D_{test} \cup \{\tau_2, \tau_3, \ldots, \tau_{3n+2}, \tau_{3n+3}\}$.

10:    $m \leftarrow m + \beta N/3$.

11:    **for** gradient step $k = 1$ to $K$ **do**

12:        Sample a batch size $N_b$ of data transitions $(s_t, a_t, r_t, s_{t+1})$ from $D_{train}$.

13:        Optimize the parameters of the actor network via **Eqs. (4), (9)**: $\theta_a \leftarrow \theta_a - \alpha_a \nabla_{\theta_a} \mathcal{J}(\theta_a)$.

14:        Optimize the parameters of the critic network via **Eqs. (3), (10)**: $\theta_c \leftarrow \theta_c - \alpha_c \nabla_{\theta_c} \mathcal{J}(\theta_c)$.

15:    **end for**

16:    Set the candidate policy: $\pi_{can} \leftarrow \pi_{\theta_a}$.

17:    Compute the importance-weighted returns $\mathcal{X}(D_{test}) = [X_1, X_2, \ldots, X_n]$ via **Eq. (11)**.

18:    Estimate lower-bound performance of candidate policy $\rho_-^{BCa}(\mathcal{X}(D_{test}), \pi_{can}, \delta)$ via **Algorithm 1**.

19:    Choose the policy $\pi^*$ via **Eq. (20)**.

20:    **if** $\pi^* = \pi_{can}$ **then**

21:        $\pi_{cur} \leftarrow \pi_{can}, \mathcal{B}_{\pi_{cur}} \leftarrow \phi, D_{train} \leftarrow \phi, D_{test} \leftarrow \phi$.

22:    **end if**

23: **end while**

---

### 2.6.2   State and action

For the automated driving planning, the state space design incorporates information about the six surrounding vehicles within a 150-meter range of the ego vehicle. These six vehicles are positioned respectively in front, rear, front-left, rear-left, front-right, and rear-right of the ego vehicle. The state information includes the relative longitudinal position, relative lateral position, and relative speed of the surrounding vehicles with respect to the ego vehicle. Additionally, the state space considers the information of the ego vehicle, including longitudinal position, lateral position and speed. Consequently, the state space is represented as a 21-dimensional vector, defined as follows:

$$S = [p_e, \{p_1, p_2, \ldots p_6\}] \tag{21}$$

with

$$\begin{aligned} p_e &= [x_e, y_e, v_e] \\ p_i &= [x_i^r, y_i^r, v_i^r] \end{aligned} \tag{22}$$

where $p_e$ denotes the ego vehicle's state, including the longitudinal position $x_e$, lateral position $y_e$ and speed $v_e$, $p_i$ represents the $i^{th}$ surrounding vehicle's relative state, including relative longitudinal position $x_i^r$, relative lateral position $y_i^r$ and relative speed $v_i^r$.

In the proposed planner, the action space involves both longitudinal and lateral actions. Besides, the action space is continuous. Consequently, the action space is represented as a 2-dimensional vector, defined as follows:

$$A = [a_{des}, \delta_{fdes}] \tag{23}$$

where $a_{des}$ is the desired acceleration or deceleration, $\delta_{fdes}$ is the desired front wheel angle.

### 2.6.3   Reward function

The reward function is designed by considering the four main aspects: driving efficiency, comfort, risk and collision. The total reward function is defined as follows:

*Jia Hu, Xuerun Yan, Tian Xu, Haoran Wang*

$$r = r_{efficiency} + r_{comfort} + r_{risk} + r_{collision} \tag{24}$$

Each aspect of the reward function is described as follows:

1) Driving efficiency

The driving efficiency is designed to encourage the ego vehicle to pursue higher speeds within the maximum speed limit. It is defined as follows:

$$r_{efficiency} = \omega_1 \frac{v_e}{v_{limit}} \tag{25}$$

where $\omega_1$ is the weight for driving efficiency, $v_{limit}$ is the maximum speed limit.

2) Comfort

The comfort reward is designed to penalize the excessive jerk and the excessive front wheel angle. It is defined as follows:

$$r_{comfort} = \omega_2 \zeta_{jerk} + \omega_3 \zeta_\delta \tag{26}$$

with

$$\zeta_{jerk} = \begin{cases} \dddot{v}_e, & if \ |\dddot{v}_e| \geq 2m/s^3 \\ 0, & else \end{cases} \tag{27}$$

$$\zeta_\delta = \begin{cases} \delta_f, & if \ |\delta_f| \geq 0.30 \ rad \\ 0, & else \end{cases} \tag{28}$$

where $\omega_2$ and $\omega_3$ are the weights for jerk and front wheel angle respectively, $\dddot{v}_e$ is vehicle jerk.

3) Risk

The risk design is to penalize the unsafe distance to the surrounding vehicles. The risk level is represented as the negative exponential function related to the time headway from the ego vehicle to the surrounding vehicle. The reward function regarding the risk is defined as follows:

$$r_{risk} = \omega_4 e^{-\frac{|x_{front}^r|}{v_e}} + \omega_5 e^{-\frac{|x_{rear}^r|}{v_{rear}}} \tag{29}$$

where $\omega_4$ and $\omega_5$ are the weights for front vehicle and rear vehicle risk respectively, the $x_{front}^r$ and $x_{rear}^r$ are the relative longitudinal position of the front and rear vehicle to the ego vehicle, $v_{rear}$ is the speed of the rear vehicle, $|x_{front}^r|/v_e$ and $|x_{rear}^r|/v_{rear}$ represent the time headway from the ego vehicle to the front vehicle and rear vehicle, respectively.

4) Collision

The collision reward function is to penalize the events where the ego vehicle collides with other vehicles or road curbs. It is designed as follows:

$$r_{collision} = \omega_6 I_c \tag{30}$$

with

$$I_c = \begin{cases} 1, & if \ collision \ happens \\ 0, & else \end{cases} \tag{31}$$

where $\omega_6$ is the weight for collision, $I_c$ is the collision indicator.

*2.6.4 Network and hyperparameter*

The actor network and critic network are both structured with four fully connected layers, including an input layer, an output layer and two hidden layers. Each hidden layer consists of 256 neurons. Tanh activation functions are applied to the hidden layers. The main hyperparameters of the proposed HCPI-RL are specified in **Table 1.**

**Table 1. Main hyperparameters of the proposed HCPI-RL**

| Parameters | Value | Parameters | Value |
|---|---|---|---|
| Maximum training step $M$ | 400000 | Number of bootstrap samples $B$ | 2000 |
| Batch size $N_b$ | 64 | Entropy coefficient $c$ | 0.01 |
| Epochs size $K$ | 10 | Weight for driving efficiency $\omega_1$ | 1.5 |
| Discounted rate $\gamma$ | 0.995 | Weight for comfort $[\omega_2, \omega_3]$ | [-0.05, -2] |
| Clipping magnitude $\epsilon$ | 0.2 | Weight for risk $[\omega_4, \omega_5]$ | [-0.5, -0.5] |
| Collected trajectories number $\beta$ | 39 | Weight for collision $\omega_6$ | -20 |
| Learning rate of actor network $\alpha_a$ | 0.0003 | Acceleration range | [-5, 2] $m/s^2$ |
| Learning rate of critic network $\alpha_c$ | 0.0003 | Front wheel angle range | [-0.7, 0.7] rad |

## 3   Performance evaluation

In this section, the performance of the proposed HCPI-RL planner is evaluated in the context of traffic. The evaluation covers the following aspects: i) monotonic evolution capability; ii) various scenarios adapting; iii) sensitivity analysis of confidence level.

### 3.1   Test bed

A previously developed joint platform combining Vissim and PreScan is utilized for the evaluations (Hu et al., 2024; Lai et al., 2020). The simulation is conducted by connecting Vissim with PreScan. The Vissim is used to build test environments and generate mixed traffic flows including human-driven vehicles and automated vehicles. The behavior of the automated vehicles is controlled through Vissim's external DriverModel and Component Object Model (COM) API. The PreScan is utilized to provide high-fidelity vehicle dynamic and actuator models, enabling realistic vehicle operations.

### 3.2   Test scenarios

Two types of test scenarios are designed for the evaluations.

**Scenario 1:** Emergent scenarios

In these scenarios, the automated vehicle is evaluated to handle the emergent cases without collision. As illustrated in **Fig. 4,** there are two cases designed: i) case 1: a truck in the adjacent lane, 15 meters ahead, suddenly cuts in front of the ego vehicle. ii) case 2: a front vehicle, 10 meters ahead, performs an emergency brake at $-8m^2$.

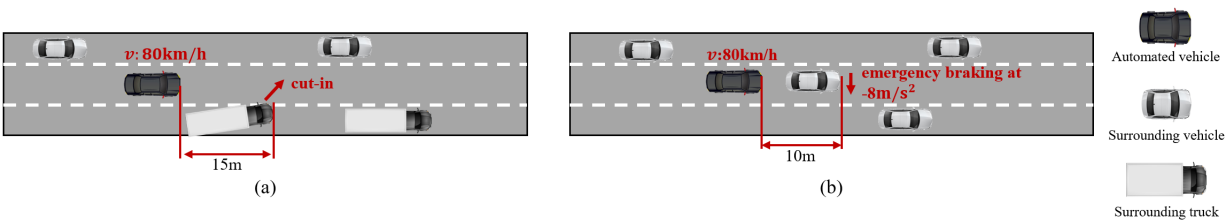

**Fig. 4.** Scenarios with emergency. (a) case 1: an adjacent truck cuts in suddenly, (b) case 2: a front vehicle conducts an emergency braking.

**Scenario 2:** Daily cruising

As depicted in **Fig. 5,** a 3-lane freeway with congested traffic is adopted to evaluate the performance of the proposed planner in the daily dynamic cruising scenario. In this scenario, an automated vehicle is cruising among the background traffic. The traffic demand is set as 2000 vehicles per lane, and the traffic desired speed distribution is set as [90, 120] km/h. During the cruising, the driving performance shall be enhanced monotonically for the proposed HCPI-RL planner.

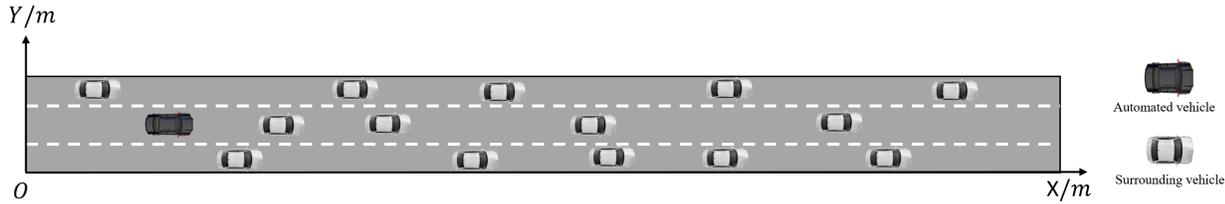

**Fig. 5.** Daily cruising scenarios with congested traffic.

### 3.3    Planner types

To benchmark the proposed HCPI-RL planner, the experiment comparisons are conducted among three kinds of planners. These planners are as follows:

- **Proposed HCPI-RL planner**: this planner is the proposed HCPI-RL planner. It updates policy with a high confidence guarantee of performance enhancement. The default confidence level is 0.90.
- **Rule-based planner**: this planner utilizes the rule-based driver model. It adopts the Mobil model (Kesting et al., 2007) as a lateral planner and the IDM model (Treiber and Kesting, 2013) as a longitudinal planner.
- **PPO planner**: this planner employs Proximal Policy Optimization (PPO) (Schulman et al., 2017) to implement RL-based automated driving.

In the online implementation of the RL-based planner, a hybrid policy is utilized to combine the rule-based planner with the RL-based planner, including the HCPI-RL and PPO planner. The rule-based method serves as a safety checker and is used when the RL-based method performs worse than the rule-based method. The algorithm of the PPO & rule-based hybrid policy is borrowed from (Cao et al., 2022; Cao et al., 2021).

### 3.4    Measurements of effectiveness

The following Measurements of Effectiveness (MOEs) are utilized for performance evaluation:

- **Return**: Cumulative reward that an agent receives from the environment over an episode of time steps. It is used to evaluate the policy's overall performance.
- **Return composition of efficiency, comfort, risk and collision**: The cumulative reward of four aspects, including efficiency, comfort, risk and collision. It is used to evaluate the policy performance in the above four aspects.
- **Success rate**: The percentage of successful driving among 100 test episodes. In the emergent scenarios, successful driving is defined as the ego vehicle driving in the emergent cases without any collisions. In the daily cruising scenarios, successful driving is characterized by the ego vehicle cruising among background traffic over 1000 meters without any collisions.
- **Vehicle trajectory**: The vehicle position at each time step.
- **Average speed**: The average speed of the ego vehicle over 1000 meters.
- **Lane change times**: The number of vehicle lane changes over 1000 meters.

### 3.5    Results

Results confirm the capability of the proposed planner in both emergent scenarios and daily cruising scenarios. They demonstrate that the proposed planner is able to achieve the aforementioned objectives: enhance the automated driving performance monotonically and adapt to the various scenarios. Specifically, the proposed HCPI-RL planner enhances the driving policy returns by 44.7% in emergent cut-in scenarios, 108.2% in emergent braking scenarios, and 64.4% in daily cruising scenarios compared to the PPO planner. Additionally, it improves the driving success rate by an average of 13.67% over the PPO planner and 61% over the rule-based planner. In the remainder of this section, more detailed results will be further discussed.

### 3.5.1 Offline training in emergent scenarios

**Fig. 6** illustrates the learning curves of HCPI-RL, PPO and rule-based planner in emergent case 1 and case 2. The figure demonstrates that the proposed HCPI-RL enables a monotonic performance enhancement of automated driving. In comparison to PPO, the proposed HCPI-RL finally receives a 44.7% and 108.2% greater return in case 1 and case 2 respectively. Moreover, the performance of HCPI-RL shows steady improvement with simulation steps, while PPO's performance curve displays volatility. It verifies the capability of HCPI-RL to guarantee a monotonic performance enhancement at a 0.90 level of confidence, while PPO cannot ensure performance enhancement at all. Furthermore, compared with the rule-based planner, the proposed HCPI-RL yields a 280% higher cumulative reward. In summary, the HCPI-RL can not only guarantee continuous performance enhancement but also achieve better performance.

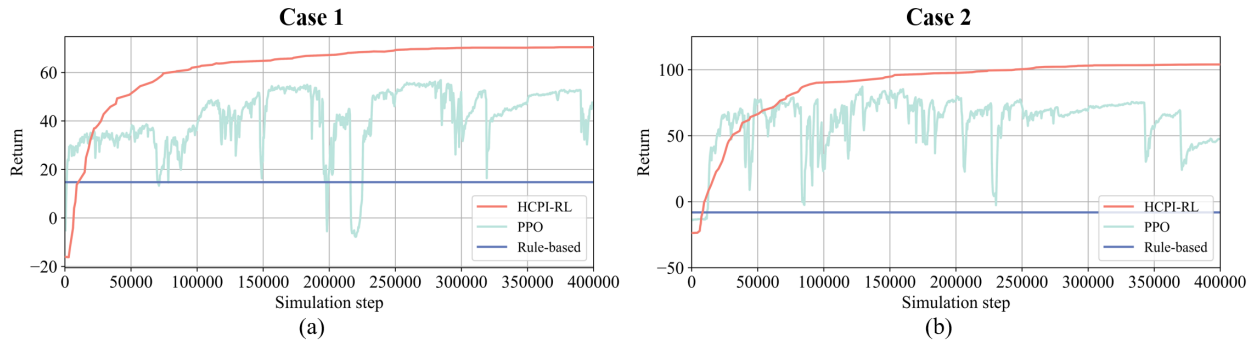

**Fig. 6.** Learning curves of HCPI-RL, PPO and rule-based planner in case 1 and case 2. (a) case 1, (b) case 2.

**Fig. 7** and **Fig. 8** depict the training process of the vehicle trajectories in case 1 and case 2, respectively. The results confirm that the proposed HCPI-RL gradually enhances performance in emergent scenarios by avoiding collisions. As shown in **Fig. 7(a)**, the ego vehicle initially collides with a cut-in truck at the beginning of the simulation. After 18,000 simulation steps of training, as shown in **Fig. 7(b)**, the ego vehicle successfully changes lanes and avoids a collision with the truck. However, after the lane change, the ego vehicle exhibits unstable lateral maneuvers, leading to fluctuations in its lateral position and a decrease in driving comfort. After nearly 200,000 simulation steps of training, as shown in **Fig. 7(c)**, the ego vehicle changes lanes and then accelerates to pursue the desired speed without unstable lateral maneuvers. Similarly, **Fig. 8** demonstrates the ego vehicle's enhanced performance in handling case 2. After around 175000 simulation steps of training, the ego vehicle overtakes the preceding vehicle smoothly, when the preceding vehicle conducts an emergency braking. In both cases, the proposed HCPI-RL finally ensures ego vehicles' driving safety and enhances driving comfort. Additionally, the proposed planner demonstrates flexibility in responding to scenario changes. As shown in **Fig. 8(b)**, the ego vehicle can adapt by changing lanes to the adjacent lane or returning to the original lane seamlessly. This flexibility is attributed to the integration of both behavior decision-making and motion planning within the proposed planner. Such an integrated design enables a timely response to environmental changes. Overall, the proposed HCPI-RL is able to handle the scenarios with emergency.

*Jia Hu, Xuerun Yan, Tian Xu, Haoran Wang*

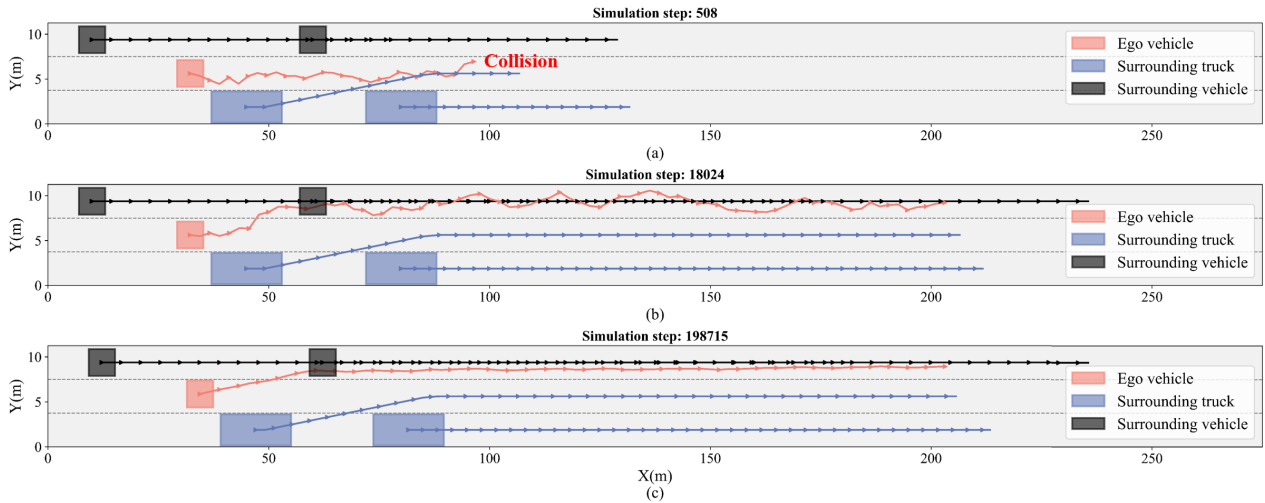

**Fig. 7.** The training process of the vehicle trajectories in case 1. (a) at 508 simulation step, (b) at 18204 simulation step, (c) at 198715 simulation step.

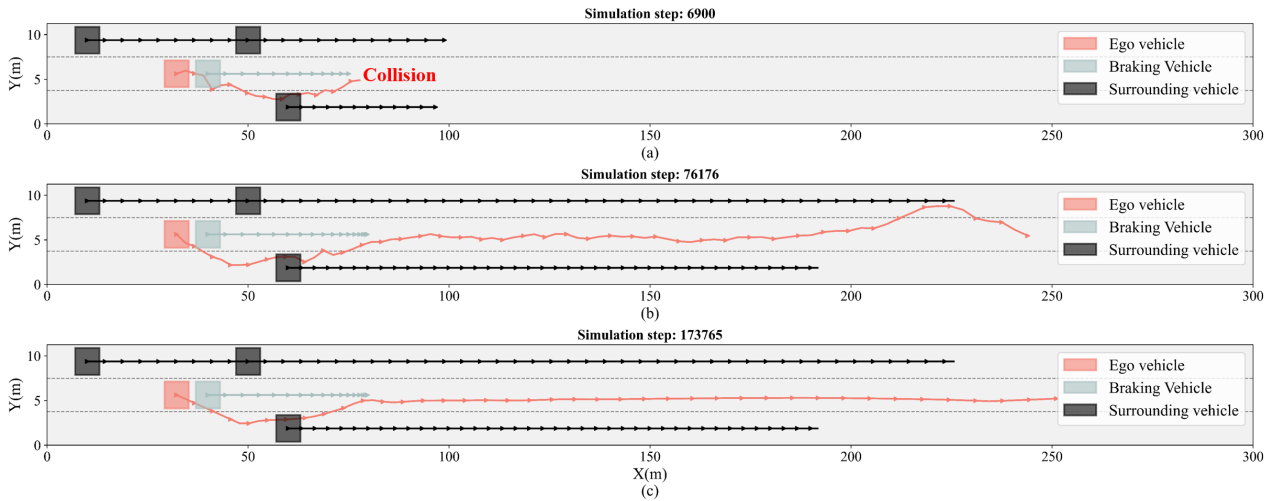

**Fig. 8.** The training process of the vehicle trajectories in case 2. (a) at 6900 simulation step, (b) at 76176 simulation step, (c) at 173765 simulation step.

**Fig. 9** and **Fig. 10** illustrate the policy returns cross four metrics in case 1 and case 2, including efficiency, comfort, risk, and collision. Three methods, HCPI-RL, PPO and rule-based, are compared. The results demonstrate the monotonic performance enhancement of HCPI-RL across the above metrics. Compared to the PPO, in case 1, HCPI-RL finally increases approximately 3.5% of driving efficiency reward and 60.3% of comfort reward. It also reduces the driving risk penalty by 28.2% and eliminates the collision penalty entirely. In case 2, HCPI-RL enhances the efficiency reward by approximately 27.7% and the comfort reward by 78.8%, while reducing the driving risk penalty by 4.2%. In contrast to rule-based planner, HCPI-RL achieves a higher efficiency reward and a lower collision penalty. However, it results in a lower comfort reward and a higher risk penalty. This makes sense as the HCPI-RL agent utilizes an aggressive policy to address the emergent scenarios successfully, whereas the rule-based vehicle adopts a conservative policy and fails in such scenarios. Furthermore, as shown in these figures, HCPI-RL shows stable enhancement of return for all four metrics, while PPO exhibits significant fluctuations. It is interesting to note that HCPI-RL initially decreases comfort and risk returns but subsequently improves these returns continuously. This initial drop is attributed to HCPI-RL agent's aggressive actions aimed at maximizing efficiency with less consideration of driving comfort and risk. Afterwards, the agent adjusts its actions to enhance comfort reward and reduce risk penalty. In summary, the HCPI-RL enables stable improvement across efficiency, comfort, risk, and collision returns.

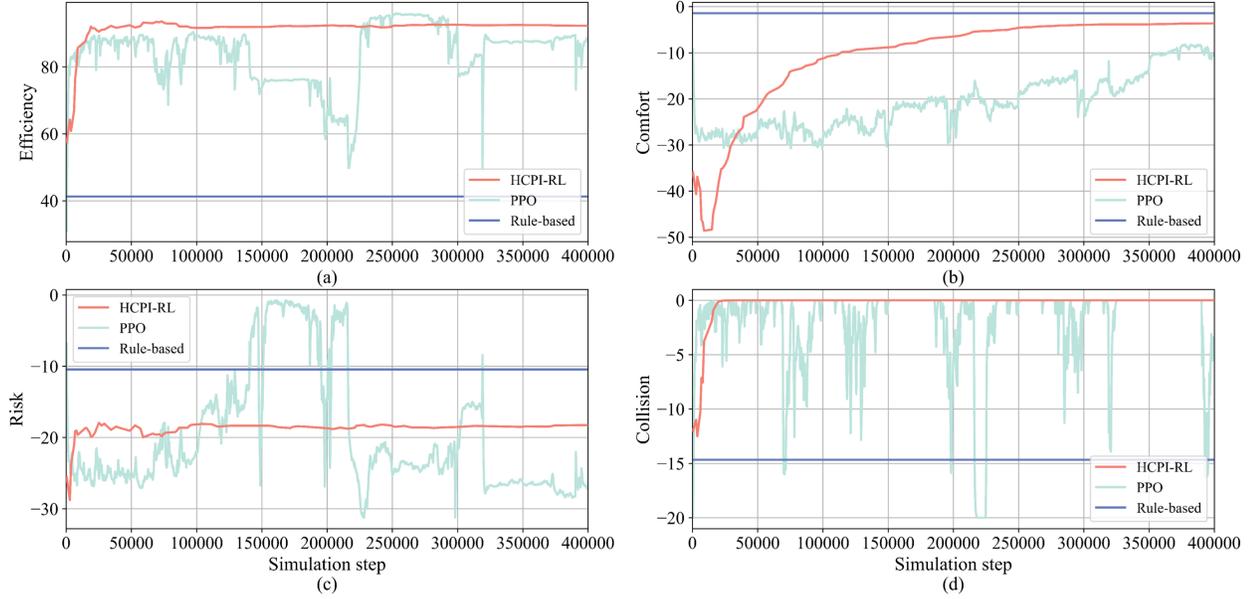

**Fig. 9.** Four aspects of performance for HCPI-RL, PPO and rule-based planner in case 1. (a) efficiency performance, (b) comfort performance, (c) risk performance, (d) collision performance.

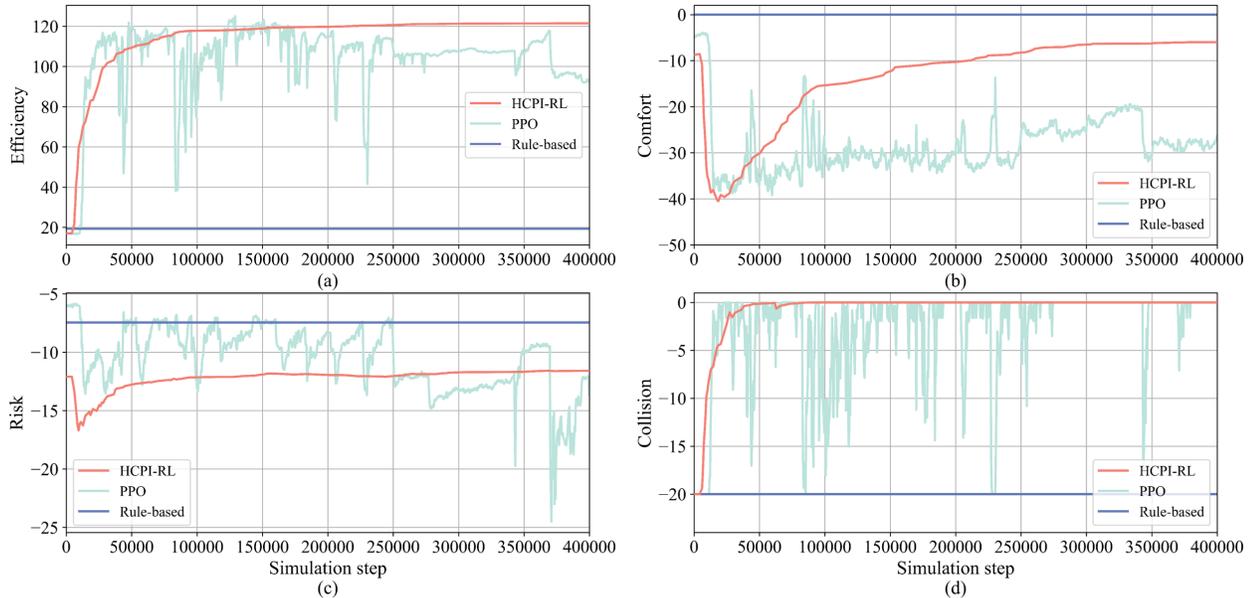

**Fig. 10.** Four aspects of performance for HCPI-RL, PPO and rule-based planner in case 2. (a) efficiency performance, (b) comfort performance, (c) risk performance, (d) collision performance.

**Fig. 11** demonstrates the success rate for the well-trained HCPI-RL, PPO and rule-based planner. The results indicate that the well-trained HCPI-RL can handle emergent scenarios with no collisions. In case 1, the HCPI-RL enhances the success rate by 20% and 73% in contrast to PPO and rule-based planners, respectively. In case 2, the HCPI-RL performs on par with PPO and improves the 100% success rate over the rule-based planner. Therefore, the results validate the HCPI-RL's functionality of successfully handling emergent scenarios.

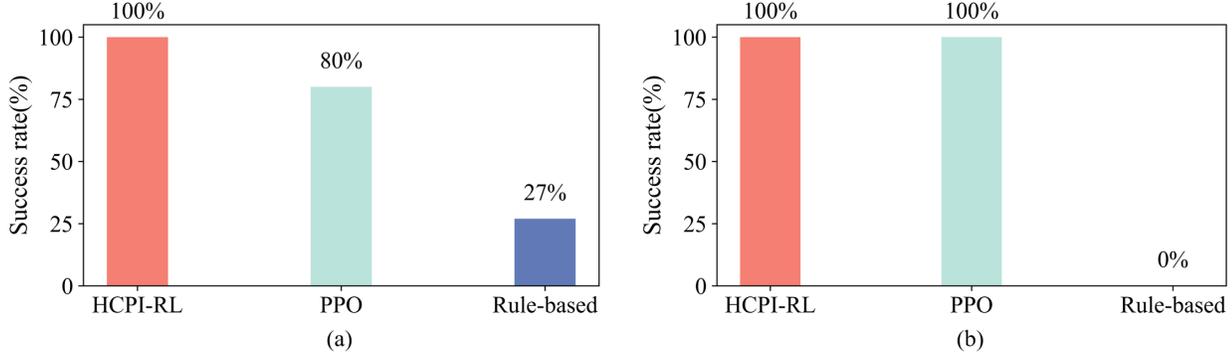

**Fig. 11.** Success rate of the automated vehicle in case 1 and case 2. (a) case 1, (b) case 2.

**Fig. 12** illustrates the return of the HCPI-RL policy at confidence levels of 0.70, 0.80, and 0.90. The results demonstrate the impact of confidence levels on HCPI-RL policy iteration. As shown in the figures, a higher confidence level results in fewer policy iterations and more stable policy performance enhancement. This indicates that the strictness of the confidence level affects the efficiency and stability of policy improvement. On the one hand, the efficiency of policy improvement decreases with a stricter confidence level. This is because the agent needs to collect more data to find a policy that meets the stricter confidence level constraints. On the other hand, the stability of policy improvement increases with a stricter confidence level. This makes sense as a stricter confidence level provides a higher guarantee of performance enhancement. Therefore, future users can dynamically adjust the confidence level to tune the trade-off between efficiency and stability of policy improvement.

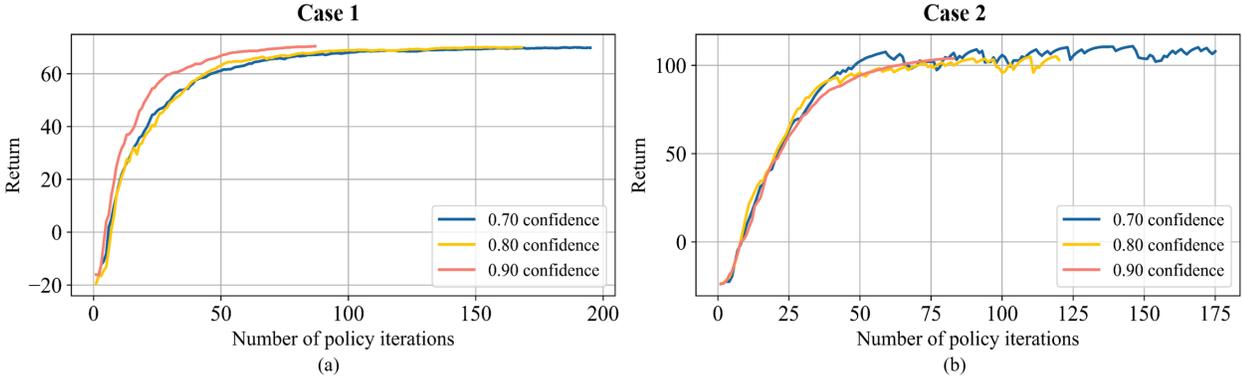

**Fig. 12.** Performance comparison among 0.70, 0.80 and 0.90 confidence levels in case 1 and case 2. (a) case 1, (b) case 2.

### 3.5.2    *Online evolving in daily cruising*

When applied in daily cursing scenarios, the performance of the proposed HCPI-RL and PPO planner is illustrated in **Fig. 13**. In the online implementation, a rule-based safety checker is adopted to ensure cruising safety. When the return of the RL policy is less than the safety checker, the RL policy would be substituted by the rule-based policy. As shown in **Fig. 13(a)**, at the early training stage, the HCPI-RL method would adopt the rule-based policy. Along with the simulation process, once the HCPI-RL surpasses the safety checker, the planner adheres to the HCPI-RL policy. This is attributed to the confidence guaranteed in the monotonic performance enhancement of each new policy. In contrast, as shown in **Fig. 13(b)**, the PPO planner frequently switches between the rule-based policy and the PPO policy due to the PPO's inconsistent performance enhancement across simulation steps. Bearing the strengths of monotonic performance enhancement, the proposed HCPI-RL planner finally achieves approximately 64.4% enhancement of return compared to the PPO planner. In summary, the above results confirm the evolution capability of the proposed HCPI-RL in daily cruising scenarios.

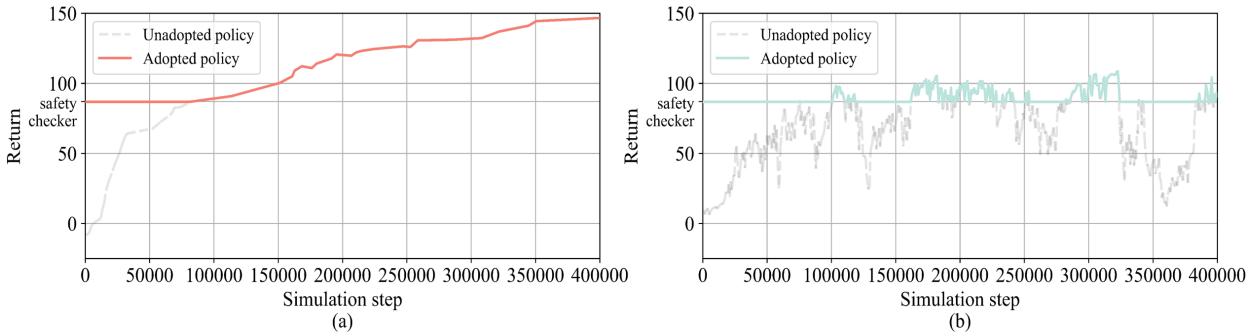

**Fig. 13.** Performance of the HCPI-RL and PPO planner. (a) hybrid policy combining the HCPI-RL and rule-based safety checker, (b) hybrid policy combining PPO and rule-based safety checker.

**Fig. 14** illustrates the success rate for the HCPI-RL, PPO and rule-based planner in daily cruising scenarios. The results demonstrate the HCPI-RL does enhance the success rate. The magnitude of enhancement is 21% and 10% compared to PPO and rule-based planners, respectively. Additionally, it is interesting that the rule-based planner outperforms the PPO planner with an 11% greater success rate. This is because daily cruising scenarios with congested traffic are complex and dynamic. The PPO method cannot adapt flexibly to great changes in these scenarios, whereas the general rule-based method can effectively cover a wide range of daily cruising scenarios. Overall, the above results confirm the HCPI-RL planner is more reliable and robust to handle the daily cruising scenarios.

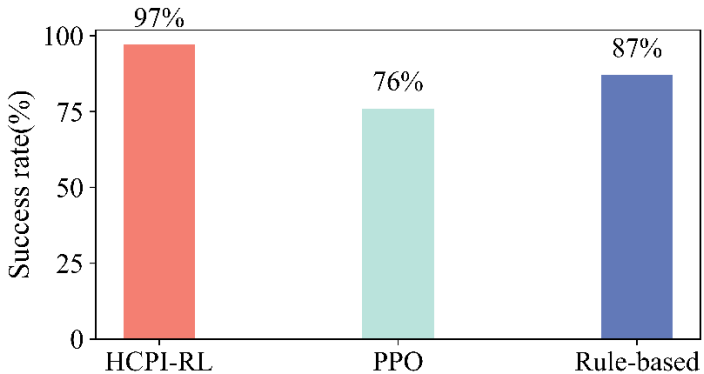

**Fig. 14.** Success rate of automated vehicle in the daily cruising scenario.

**Fig. 15** illustrates the average speed and lane change times for the HCPI-RL, PPO and rule-based planner in daily cruising scenarios. The results demonstrate the efficiency of the HCPI-RL planner. The HCPI-RL method obtains an approximately 19.2% and 30.7% greater average speed compared to PPO and rule-based planners, respectively. Additionally, the HCPI-RL enhances approximately 45.8% average lane change times compared to PPO and 250% compared to rule-based planner. These results reveal that the HCPI-RL planner is able to compete with the surrounding vehicles and change lanes to maintain an efficient speed in congested traffic. To further validate its functionality of cutting through congested traffic, the trajectory is depicted for HCPI-RL as shown in **Fig. 16**. In this figure, the HCPI-RL method enables the ego vehicle to change lanes frequently and overtake 6 slow-moving vehicles within 15 seconds. Such an overtake maneuver enhances driving efficiency in congested traffic conditions.

*Jia Hu, Xuerun Yan, Tian Xu, Haoran Wang*

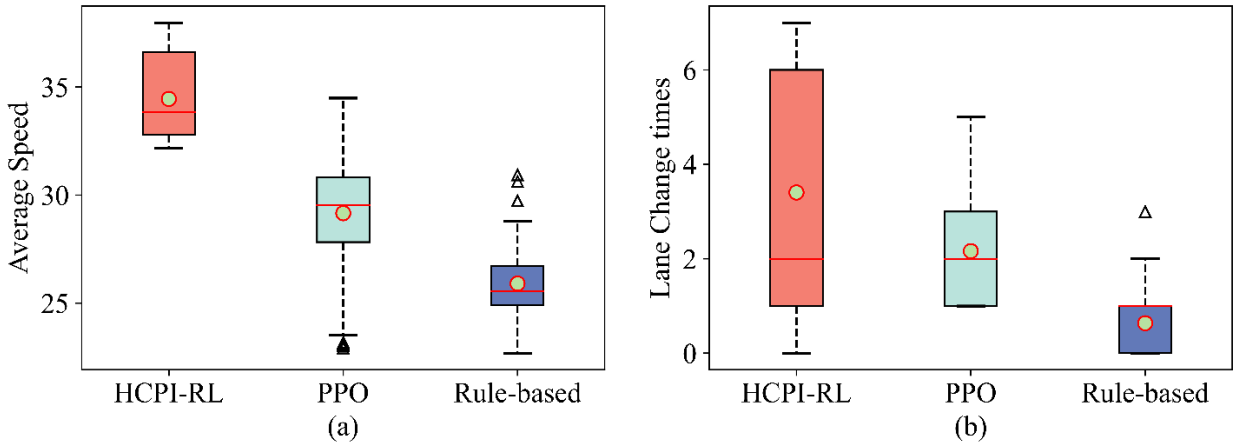

**Fig. 15.** Average speed and lane change times of automated vehicle. (a) average speed, (b) lane change times.

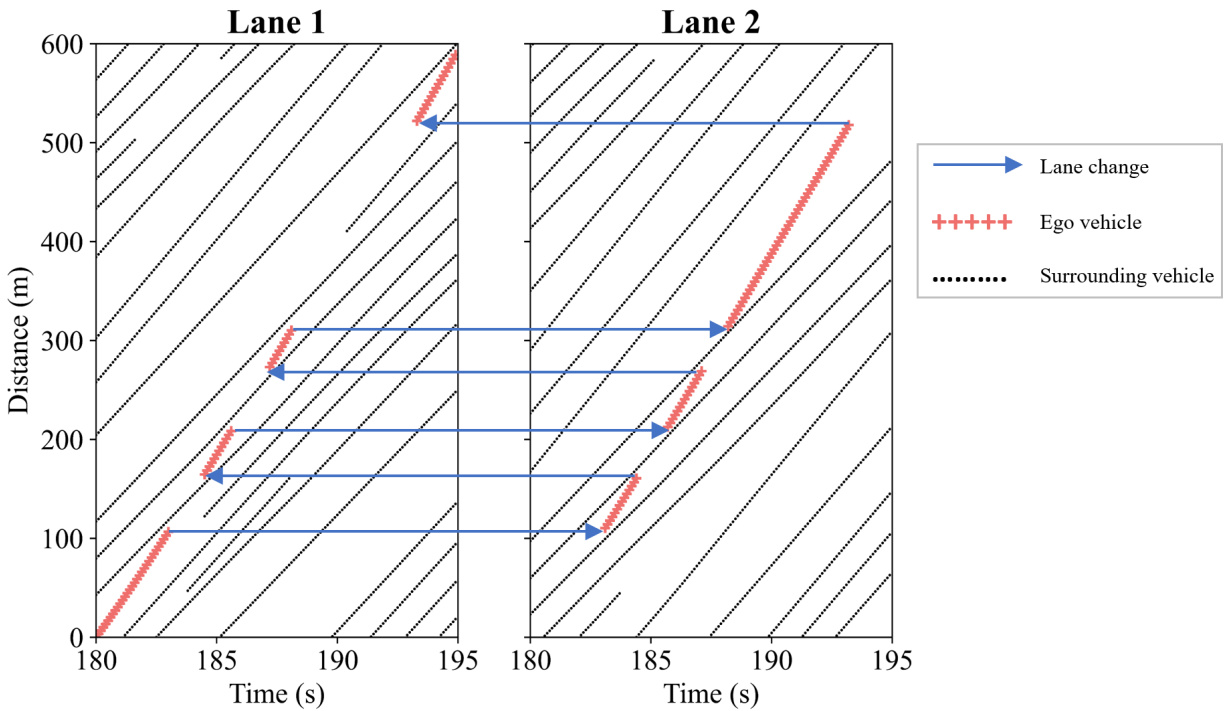

**Fig. 16.** Vehicle trajectories using the HCPI-RL policy.

## 4 Conclusion

This research proposes a High Confidence Policy Improvement Reinforcement Learning-based (HCPI-RL) planner. It is designed to achieve the monotonic evolution of automated driving. The proposed HCPI-RL planner has the following features: i) Evolutionary automated driving with monotonic performance enhancement; ii) With the capability of handling scenarios with emergency; iii) With enhanced decision-making optimality. A novel policy update paradigm is designed to enable the newly learned policy's performance always surpasses that of previous policies, which is deemed as monotonic performance enhancement. The proposed planner is evaluated on a joint platform that consists of Vissim and PreScan. The proposed planner's performance is quantified in both emergent scenarios and daily cruising scenarios. Results demonstrate that:

- The proposed planner functions as expected and is able to enhance the automated driving performance monotonically.
- The trade-off between policy improvement efficiency and stability is able to be tuned by adjusting the confidence level of policy update.
- In contrast to the conventional PPO planner, the proposed planner enhances performance due to greater driving policy returns by 44.7% in emergent cut-in scenarios, 108.2% in emergent braking scenarios, and 64.4% in daily cruising scenarios.
- Compared to the PPO planner, the proposed planner achieves a higher performance cross efficiency, comfort, risk and collision. The proposed planner increases approximately 15.6% of driving efficiency reward and 69.6% of comfort reward. It also reduces 16.2% of driving risk penalty and 50% of collision penalty.
- The proposed planner is confirmed to enhance the driving success rate in both emergent scenarios and daily cruising scenarios. On average, it reduces collisions by 13.67% compared to the PPO planner and by 61% compared to the rule-based planner.
- The proposed planner enhances driving efficiency in congested daily cruising scenarios. It obtains an approximately 19.2% greater average speed and 45.8% more average lane change times compared to the PPO planner, and approximately 30.7% greater average speed and 250% more average lane change times compared to the rule-based planner.

In this research, the proposed HCPI-RL planner does not consider driver preference. Further study could enhance the proposed method through designing a personalized reward function. Such an enhancement could enable the automated driving consistently approach to personalized driver behaviors. Additionally, this research focuses on freeway scenarios. In future research, more completed scenarios, such as intersections and conflicts with pedestrians, should be explored to further validate the proposed method's capability.

**Acknowledgments**

This paper is partially supported by National Key R&D Program of China (2022YFE0117100), National Natural Science Foundation of China (Grant No. 52302412 and 52372317), Yangtze River Delta Science and Technology Innovation Joint Force (No. 2023CSJGG0800), Shanghai Automotive Industry Science and Technology Development Foundation (No. 2404), Xiaomi Young Talents Program, the Fundamental Research Funds for the Central Universities, Tongji Zhongte Chair Professor Foundation (No. 000000375-2018082), the Postdoctoral Fellowship Program (Grade B) of China Postdoctoral Science Foundation (GZB20230519), Shanghai Sailing Program (No. 23YF1449600), Shanghai Post-doctoral Excellence Program (No.2022571), China Postdoctoral Science Foundation (No.2022M722405), and the Science Fund of State Key Laboratory of Advanced Design and Manufacturing Technology for Vehicle (No. 32215011).

*Jia Hu, Xuerun Yan, Tian Xu, Haoran Wang*

# References

Abbeel, P., Ng, A.Y., 2004. Apprenticeship learning via inverse reinforcement learning, *Proceedings of the twenty-first international conference on Machine learning*, p. 1.

Al-Sharman, M., Dempster, R., Daoud, M.A., Nasr, M., Rayside, D., Melek, W., 2023. Self-learned autonomous driving at unsignalized intersections: A hierarchical reinforced learning approach for feasible decision-making. *IEEE Transactions on Intelligent Transportation Systems* 24(11), 12345-12356.

Altman, E., 2021. *Constrained Markov decision processes*. Routledge.

Aradi, S., 2020. Survey of deep reinforcement learning for motion planning of autonomous vehicles. *IEEE Transactions on Intelligent Transportation Systems* 23(2), 740-759.

Baheri, A., Nageshrao, S., Tseng, H.E., Kolmanovsky, I., Girard, A., Filev, D., 2020. Deep reinforcement learning with enhanced safety for autonomous highway driving, *2020 IEEE Intelligent Vehicles Symposium (IV)*. IEEE, pp. 1550-1555.

Cao, Z., Jiang, K., Zhou, W., Xu, S., Peng, H., Yang, D., 2023. Continuous improvement of self-driving cars using dynamic confidence-aware reinforcement learning. *Nature Machine Intelligence* 5(2), 145-158.

Cao, Z., Xu, S., Jiao, X., Peng, H., Yang, D., 2022. Trustworthy safety improvement for autonomous driving using reinforcement learning. *Transportation research part C: emerging technologies* 138, 103656.

Cao, Z., Xu, S., Peng, H., Yang, D., Zidek, R., 2021. Confidence-aware reinforcement learning for self-driving cars. *IEEE Transactions on Intelligent Transportation Systems* 23(7), 7419-7430.

Changxi You, J.L., Dimitar Filev, Panagiotis Tsiotras, 2019. Advanced planning for autonomous vehicles using reinforcement learning and deep inverse reinforcement learning. *Robotics and Autonomous Systems* Volume 114, Pages 1-18.

Chen, D., Hajidavalloo, M.R., Li, Z., Chen, K., Wang, Y., Jiang, L., Wang, Y., 2023. Deep multi-agent reinforcement learning for highway on-ramp merging in mixed traffic. *IEEE Transactions on Intelligent Transportation Systems* 24(11), 11623-11638.

Du, J., Bai, Y., Li, Y., Geng, J., Huang, Y., Chen, H., 2024. Evolutionary End-to-End Autonomous Driving Model With Continuous-Time Neural Networks. *IEEE/ASME Transactions on Mechatronics*.

Elallid, B.B., Benamar, N., Hafid, A.S., Rachidi, T., Mrani, N., 2022. A comprehensive survey on the application of deep and reinforcement learning approaches in autonomous driving. *Journal of King Saud University-Computer and Information Sciences* 34(9), 7366-7390.

Gu, Z., Gao, L., Ma, H., Li, S.E., Zheng, S., Jing, W., Chen, J., 2023. Safe-state enhancement method for autonomous driving via direct hierarchical reinforcement learning. *IEEE Transactions on Intelligent Transportation Systems* 24(9), 9966-9983.

He, X., Hao, J., Chen, X., Wang, J., Ji, X., Lv, C., 2024a. Robust Multiobjective Reinforcement Learning Considering Environmental Uncertainties. *IEEE Transactions on Neural Networks and Learning Systems*.

He, X., Huang, W., Lv, C., 2024b. Trustworthy autonomous driving via defense-aware robust reinforcement learning against worst-case observational perturbations. *Transportation Research Part C: Emerging Technologies* 163, 104632.

Hu, J., Yan, X., Wang, G., Tu, M., Zhang, X., Wang, H., Gruyer, D., Lai, J., 2024. A Simulation Platform for Truck Platooning Evaluation in an Interactive Traffic Environment. *IEEE Transactions on Intelligent Transportation Systems*.

Huang, Y., Yang, S., Wang, L., Yuan, K., Zheng, H., Chen, H., 2023. An efficient self-evolution method of autonomous driving for any given algorithm. *IEEE Transactions on Intelligent Transportation Systems*.

Isele, D., Rahimi, R., Cosgun, A., Subramanian, K., Fujimura, K., 2018. Navigating occluded intersections with autonomous vehicles using deep reinforcement learning, *2018 IEEE international conference on robotics and automation (ICRA)*. IEEE, pp. 2034-2039.

Jung, K., Lee, J., Gupta, V., Cho, G., 2019. Comparison of bootstrap confidence interval methods for GSCA using a Monte Carlo simulation. *Frontiers in psychology* 10, 2215.

Kalra, N., Paddock, S.M., 2016. Driving to safety: How many miles of driving would it take to demonstrate autonomous vehicle reliability? *Transportation Research Part A: Policy and Practice* 94, 182-193.

Kesting, A., Treiber, M., Helbing, D., 2007. General lane-changing model MOBIL for car-following models. *Transportation Research Record* 1999(1), 86-94.

Kiran, B.R., Sobh, I., Talpaert, V., Mannion, P., Al Sallab, A.A., Yogamani, S., Pérez, P., 2021. Deep reinforcement learning for autonomous driving: A survey. *IEEE Transactions on Intelligent Transportation Systems* 23(6), 4909-4926.

Konda, V., Tsitsiklis, J., 1999. Actor-critic algorithms. *Advances in neural information processing systems* 12.

Lai, J., Hu, J., Cui, L., Chen, Z., Yang, X., 2020. A generic simulation platform for cooperative adaptive cruise control under partially connected and automated environment. *Transportation Research Part C: Emerging Technologies* 121, 102874.

Laroche, R., Trichelair, P., Des Combes, R.T., 2019. Safe policy improvement with baseline bootstrapping, *International conference on machine learning*. PMLR, pp. 3652-3661.

Le Mero, L., Yi, D., Dianati, M., Mouzakitis, A., 2022. A survey on imitation learning techniques for end-to-end autonomous vehicles. *IEEE Transactions on Intelligent Transportation Systems* 23(9), 14128-14147.

Li, Q., Peng, Z., Zhou, B., 2022. Efficient learning of safe driving policy via human-ai copilot optimization. *arXiv preprint arXiv:2202.10341*.

Li, S.E., 2023. Deep reinforcement learning, *Reinforcement learning for sequential decision and optimal control*. Springer, pp. 365-402.

Li, Z., Hu, J., Leng, B., Xiong, L., Fu, Z., 2023. An Integrated Decision Making and Motion Planning Framework for Enhanced Oscillation-Free Capability. *IEEE Transactions on Intelligent Transportation Systems*.

Lin, Y., McPhee, J., Azad, N.L., 2022. Co-optimization of on-ramp merging and plug-in hybrid electric vehicle power split using deep reinforcement learning. *IEEE Transactions on Vehicular Technology* 71(7), 6958-6968.

Lu, C., Lu, H., Chen, D., Wang, H., Li, P., Gong, J., 2023. Human-like decision making for lane change based on the cognitive map and hierarchical reinforcement learning. *Transportation research part C: emerging technologies* 156, 104328.

Metelli, A.M., Pirotta, M., Calandriello, D., Restelli, M., 2021. Safe policy iteration: A monotonically improving approximate policy iteration approach. *Journal of Machine Learning Research* 22(97), 1-83.

Nageshrao, S., Tseng, H.E., Filev, D., 2019. Autonomous highway driving using deep reinforcement learning, *2019 IEEE International Conference on Systems, Man and Cybernetics (SMC)*. IEEE, pp. 2326-2331.

Pateria, S., Subagdja, B., Tan, A.-h., Quek, C., 2021. Hierarchical reinforcement learning: A comprehensive survey. *ACM Computing Surveys (CSUR)* 54(5), 1-35.

Pirotta, M., Restelli, M., Pecorino, A., Calandriello, D., 2013. Safe policy iteration, *International conference on machine learning*. PMLR, pp. 307-315.

Schulman, J., Wolski, F., Dhariwal, P., Radford, A., Klimov, O., 2017. Proximal policy optimization algorithms. *arXiv preprint arXiv:1707.06347*.

Sutton, R.S., Barto, A.G., 2018. *Reinforcement learning: An introduction*. MIT press.

Thomas, P., Brunskill, E., 2016. Data-efficient off-policy policy evaluation for reinforcement learning, *International Conference on Machine Learning*. PMLR, pp. 2139-2148.

Thomas, P., Theocharous, G., Ghavamzadeh, M., 2015a. High-confidence off-policy evaluation, *Proceedings of the AAAI Conference on Artificial Intelligence*.

Thomas, P., Theocharous, G., Ghavamzadeh, M., 2015b. High confidence policy improvement, *International Conference on Machine Learning*. PMLR, pp. 2380-2388.

Tokdar, S.T., Kass, R.E., 2010. Importance sampling: a review. *Wiley Interdisciplinary Reviews: Computational Statistics* 2(1), 54-60.

Treiber, M., Kesting, A., 2013. Car-following models based on driving strategies, *Traffic flow dynamics*. Springer, pp. 181-204.

Van Brummelen, J., O'brien, M., Gruyer, D., Najjaran, H., 2018. Autonomous vehicle perception: The technology of today and tomorrow. *Transportation research part C: emerging technologies* 89, 384-406.

Wachi, A., Sui, Y., 2020. Safe reinforcement learning in constrained markov decision processes, *International Conference on Machine Learning*. PMLR, pp. 9797-9806.

Wu, J., Huang, C., Huang, H., Lv, C., Wang, Y., Wang, F.-Y., 2024. Recent advances in reinforcement learning-based autonomous driving behavior planning: A survey. *Transportation Research Part C: Emerging Technologies* 164, 104654.

Yang, K., Tang, X., Qiu, S., Jin, S., Wei, Z., Wang, H., 2023. Towards robust decision-making for autonomous driving on highway. *IEEE Transactions on Vehicular Technology* 72(9), 11251-11263.

Yang, Z., Zheng, Z., Kim, J., Rakha, H., 2024. Eco-driving strategies using reinforcement learning for mixed traffic in the vicinity of signalized intersections. *Transportation Research Part C: Emerging Technologies* 165, 104683.

Yuan, K., Huang, Y., Yang, S., Wu, M., Cao, D., Chen, Q., Chen, H., 2024. Evolutionary Decision-Making and Planning for Autonomous Driving: A Hybrid Augmented Intelligence Framework. *IEEE Transactions on Intelligent Transportation Systems*.

Zhang, Z., Liu, H., Lei, M., Yan, X., Wang, M., Hu, J., 2023a. Review on the impacts of cooperative automated driving on transportation and environment. *Transportation Research Part D: Transport and Environment* 115, 103607.

Zhang, Z., Yan, X., Wang, H., Ding, C., Xiong, L., Hu, J., 2023b. No more road bullying: an integrated behavioral and motion planner with proactive right-of-way acquisition capability. *Transportation research part C: emerging technologies* 156, 104363.

Zhu, Z., Zhao, H., 2021. A survey of deep RL and IL for autonomous driving policy learning. *IEEE Transactions on Intelligent Transportation Systems* 23(9), 14043-14065.